\documentclass[aps,prl,twocolumn,groupedaddress,showpacs]{revtex4}
\usepackage{graphicx}

\begin{document}

\preprint{}

\title{Hybrid 2D surface trap for quantum simulation}


\author{J.I. Gillen,  W. S. Bakr, A. Peng, P. Unterwaditzer,  S. F\"olling, M. Greiner }
\affiliation{Department of Physics and Harvard-MIT Center for
Ultracold Atoms, Harvard University, Cambridge, Massachusetts 02138, USA}



\date{\today}

\begin{abstract}
We demonstrate a novel optical trapping scheme for ultracold atoms.
Using a combination of evanescent wave, standing wave, and magnetic
potentials we create a deeply 2D Bose-Einstein condensate (BEC) at a
few microns from a glass surface. Using techniques such as
broadband ``white'' light to create evanescent and standing
waves, we realize a smooth potential with a trap frequency aspect ratio of 300:1:1 and long lifetimes. This makes the setup
suitable for many-body quantum simulations and applications such as
high precision measurements close to surfaces.
\end{abstract}


\maketitle


Ultracold quantum gases are increasingly used to experimentally
realize and quantitatively study fundamental models of condensed
matter physics \cite{Bloch08}. Two dimensional (2D) quantum
gases \cite{Gorlitz01,Rychtarik04,Stock05,Smith05,Kohl05} are systems uniquely
suited for carrying out  ``quantum simulations'' of condensed matter
phenomena, such as the Berezinskii-Kosterlitz-Thouless (BKT)
transition \cite{Kosterlitz73} recently observed in ultracold atoms
\cite{Hadzibabic06, Schweikhard07, Clade08}. These systems are
characterized by a strong confinement in one direction that sets an
energy scale much larger than that given by either the temperature or
interparticle interactions. 2D systems can be brought into the
strongly correlated regime, for example by adding lattice potentials
in the plane\cite{Kohl05,Spielman07}. Simulations in this regime
would enable studying a variety of quantum phases including 2D
antiferromagnets and d-wave superfluid states \cite{Lee06}.

The centerpiece of an experimental system for 2D quantum simulation
is an atom trap that fulfills a number of criteria. The axial confinement of the trap must
be very strong to be deep in the 2D regime even for large atom numbers, while the weak
lateral confinement should be free of disorder. Additionally, it is desirable to have a trap
configuration that enables the reliable preparation of a single plane of
atoms to avoid averaging effects during quantum state readout, as
well as dual planes for heterodyne phase measurements\cite{Stock05}.
Finally, high resolution optical access to the quantum gas would be useful for probing and
manipulation at small length scales.

\begin{figure}[htbp]
\begin{center}
\includegraphics[width=3.2in]{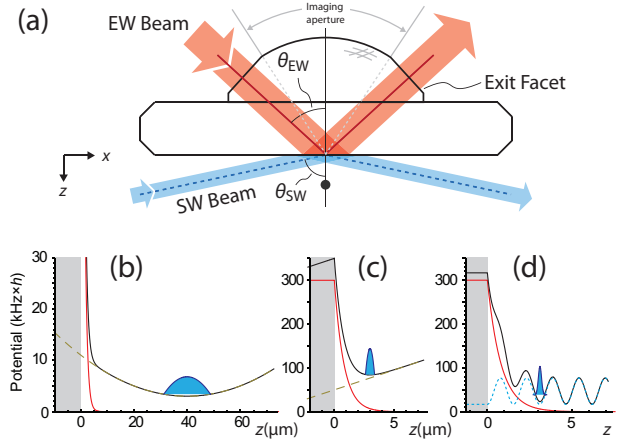}
\caption{Hybrid trap (a): The trap is based on a dielectric surface
formed by a superpolished horizontal bottom face of a fused silica
substrate. Fused to the top of the substrate is a hemispherical lens
with lateral planar facets which is part of a high numerical aperture
imaging system. Three potentials form the trap: An evanescent wave
beam (EW) from the top generates an exponential repulsive potential
(red lines). A beam reflected from the other side of the
surface creates a standing wave (SW, blue dotted lines). A parabolic
magnetic trap (green dashed lines) provides an upwards
force and lateral confinement. (b) Magnetic trap with BEC closely
below surface. The black line denotes the combined potential. (c)
The evanescent wave trap is loaded by shifting the magnetic trap
minimum inside the glass, generating confinement between the
magnetic gradient and the evanescent wave. (d) Atoms loaded into
single plane of SW trap, 3 micron from the surface. The trap
provides deep 2D confinement with typical lateral trap frequencies
of $2\pi\times 20$\,Hz and a vertical confinement of $2\pi\times
6$\,kHz.} \label{figure1}
\end{center}
\end{figure}

Different approaches for reaching the 2D regime have been experimentally
realized \cite{Bloch08}. Repulsive evanescent wave (EW) potentials were originally introduced to reflect
atoms from surfaces \cite{Balykin88,Landragin96a}. They were combined with gravity to form a
gravito-optical surface trap \cite{Ovchinnikov97,Hammes03} in which a 2D BEC was observed \cite{Rychtarik04}.
However, such traps have been limited by potential corrugations which
led to variations of trap characteristics across the surface
as well as inhibited expansion of the BEC \cite{Rychtarik04,Rychtarik04b}. The strongest axial confinement in 2D systems has been achieved in optical dipole standing wave (SW) traps where a stack of 2D planes is populated by a BEC \cite{Stock05,Kohl05,Spielman07}. It is possible to empty all but
a few planes by a radio-frequency ``knife'' \cite{Stock05}, but precisely
controlling the population remains a challenge, in particular for
traps with strong confinement and small periodicity.

Here we demonstrate a novel trapping scheme that overcomes
limitations of current methods and achieves all the
requirements stated above. Our hybrid trapping potential is based on
a dielectric surface used to combine an evanescent wave, nodal planes
of reflected light and a magnetic confinement (Fig. 1). A surface
trap approach provides us with several important advantages. Using a
high quality dielectric surface as our reference, we are able to
precisely and reproducibly overlap optical potentials. In contrast to
free-space lattices the surface approach allows us to directly load
all atoms into one or two planes. The proximity of these planes to
the surface facilitates imaging them with high resolution  due to a
solid immersion effect. This effect allows us to achieve a very high numerical aperture NA=0.8, which should allow for an imaging resolution of $\approx 0.5$ micron. The vicinity
of the gas to the surface also opens the door to precision studies of surface
physics\cite{Landragin96b,Lin04,Harber05}. By using new techniques
such as "white" light broadband dipole potentials we are able to
avoid potential corrugations. Long-lived BECs deep in the 2D
regime are obtained and we are able to achieve low temperatures by evaporative cooling in the 2D trap. We observe long wavelength
phase fluctuations, and at higher temperatures, vortices, characteristic of BKT physics.

The evanescent wave potential is created using blue detuned light that is totally
internally reflected at the interface of a dielectric medium and a
vacuum \cite{Grimm00}. The EW that appears on the
vacuum side creates an exponentially decaying potential $V_{EW}(z) =
V_0 \exp(-2z/\Lambda)$, where $z$ is the distance to the surface,
$\Lambda$ the decay length of the EW and $V_0$ the
potential height at the surface given by the total incoming
intensity.

Evanescent wave traps are particularly susceptible to disorder in the lateral potential. One very significant source of disorder is stray light that interferes with the trap light. This can cause sizable potential modulations at high spatial frequencies even for weak stray light intensities.
We address this effect in multiple ways. First, we disable the
interference between the trap light and most of the stray light from
reflections, dust and surface imperfections by using ``white'' light with
a very short coherence length. It is created by a tapered amplifier system seeded with light
from a fiber coupled amplified spontaneous emission (ASE) source.
Interference filters are used to control the bandwidth of the white light source as well as to suppress
resonant ASE components. The resulting spectrum is centered around
a wavelength of $\lambda_{trap}=765$\,nm, with an incident power of 200\,mW. The spectral width of
3\,nm corresponds to a coherence length of about 100\,$\mu$m, which
is short enough to suppress the interference effects of stray light
from multiple reflections and other parts within the setup.

A second measure is the minimization of scatter at the trapping surface itself by using a superpolished and very clean surface with an RMS surface roughness below 1\,\AA. This reduces corrugations due to scatter from locations within the range of the coherence length. We measured the residual scattered fraction at the glass vacuum interface to be less than 1\,ppm.

Neutral atoms in close vicinity to surfaces can also be affected by static electric potentials. Adsorbed metal
atoms on the surface form small electric dipoles. Inhomogeneities of the distribution of these dipoles generate potential gradients which can be stronger than the inherent van der Waals force of the substrate. This process is well understood
\cite{McGuirk04} and the fields generated decay very rapidly away from the surface. By avoiding exposure of the glass to significant amounts of rubidium and choosing a moderate $3\,\mu$m distance from the surface, detrimental effects are avoided even after months of operation.

Our experiments start with a 3D BEC of several $10^5$ $^{87}$Rb atoms. The BEC is created by loading a magneto-optic trap for 8 seconds, resulting in $10^9$ atoms at a temperature of 40\,$\mu$k after optical molasses cooling.  We then transport the atoms into a glass cell with very good vacuum \cite{Greiner01}, inside which the substrate is mounted.
Here, we perform forced RF evaporation in a tightly confining QUIC trap \cite{Esslinger98} with frequencies of $\omega_{x,y,z}=2\pi (17, 100, 100)$\,Hz to achieve BEC. Following this, we change to a high offset field configuration with reduced trap frequencies of $2\pi (17, 20, 20)$\,Hz. At this stage, the BEC is 0.6\,mm below the surface.

\begin{figure}[bthp]
\begin{center}
\includegraphics[width=3.3in]{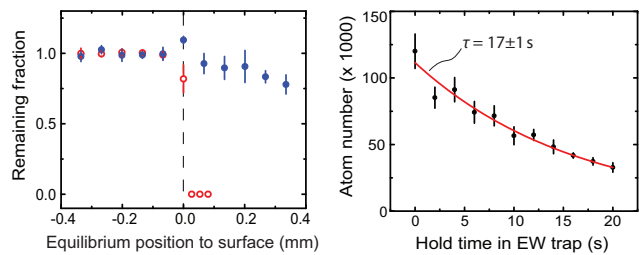}
\caption{(a) Moving the atoms against the surface without EW leads to a sudden loss (red circles), which is avoided by the repulsive potential (blue filled circles). (b) Atom lifetime in the hybrid EW trap, red line corresponds to fit with single exponential.}\label{figure2}
\end{center}
\end{figure}

The evanescent wave trap configuration is formed by moving the magnetic trap vertically upwards using external magnetic fields.
The equilibrium position is moved by $1.6$\,mm, well past the fused silica surface. When it passes the surface, the atoms are vertically held in place by the EW, giving rise to a combined opto-magnetic surface trap (Fig. \ref{figure2}a).
The EW beam is an elliptic beam of $250\times180$\,$\mu$m$^2$ size incident at an angle $\theta_{EW}$, 12\,mrad from the critical angle $\theta_c$ (Figure \ref{figure1}). The decay length $\Lambda$ is given by
 ${\lambda_{trap}}/{2\pi}/{({n^{2}\sin^{2}(\theta_{EW})-1})^{1/2}} \approx800$\,nm where $n$ is the index of refraction of fused silica. The EW potential has a maximum at a distance of 200\,nm from the surface, below which the attractive van der Waals potential dominates. 
The short decay length of the EW gives rise to large curvatures that allow tight confinement along the direction of the decay. Trap frequencies of up to $2\pi\times 1$\,kHz, measured by parametric excitation, can be reached in this configuration.  In this 2D configuration, we observe a $1/e$ atom lifetime of $17\pm 1$\,s (Fig. \ref{figure2}b). To verify the lateral homogeneity of the resulting potential, we dynamically transport the cloud over the diameter of the EW beam (by displacing the magnetic field center up to 250\,$\mu$m) without causing significant heating or atom loss due to ``holes'' in the evanescent wave.

We have implemented two methods to cool the atom cloud via forced evaporation in the surface trap.
Moving the atoms against the surface is typically not fully adiabatic and the temperature of the cloud increases. A second evaporation step enables us to regain a high condensate fraction in the surface trap.
One available method is lowering the EW power until the thermal atoms can tunnel to the glass surface. This generates a very efficient evaporation effect, but after many repetitions the dipole forces generated by adsorbed Rb atoms affect the barrier height. The diffusion timescale of these adsorbates has been shown in previous experiments to be on the scale of days to weeks and can be strongly decreased by an increase in surface temperature \cite{Obrecht07}. A second, more stable evaporation technique is to use forced RF evaporation which will be discussed later.

To further increase the vertical confinement and improve the lateral homogeneity, we use an additional standing wave potential.
The larger trap frequencies allow us to reach the deep 2D regime for higher atom numbers as well as to achieve stronger interactions in the system. The standing wave potential is generated by reflecting a blue-detuned beam off the glass surface from the vacuum side \cite{Engeser06}. Incident at an angle $\theta_{SW}=76^\circ$ from the normal, the trap minima in the resulting potential are planes parallel to the surface with a trap frequency increased by a factor of 5 or more. Contrary to all current standing wave traps, the potential is formed not by coherent light but by light from our 765\,nm broadband source to reduce disorder from stray light interference. As the atoms are very close to the reflecting surface, the remaining coherence length of the broadband light is still larger than the interfering distance $2d /\cos\theta_{SW}$.

The SW configuration also reduces further the disorder caused by remaining scattered light interference from those parts of the glass which are closer than the coherence length. As the SW has an intensity minimum at the surface as opposed to the intensity maximum of the EW, scattering from small surface impurities is suppressed.


We reliably load all of the atoms into a single node of the standing wave, as the spread of the wavefunction ($z_{ho}=250$\,nm) in the pure EW trap is much smaller than the spacing of the SW planes. The transfer is realized by smoothly transferring power from the EW beam to the standing wave beam over a period of 300\,ms.
The trap frequency in the $z$ direction, verified by parametric excitation measurements, is increased to $5.9\pm 0.1$\,kHz in this trap, taking us deep into the 2D regime with the temperature and chemical potential both much smaller than the vertical trap frequency \cite{Gorlitz01}. 
We populate the second node of the SW at a distance of $\approx\,3$\,$\mu$m  from the surface. In this plane, the disorder caused by surface dipoles is already strongly reduced, but the distance from the surface is small enough to exploit the NA enhancement when imaging from the top.
The SW trap configuration is smooth enough that the
atoms leave it by moving out of the beam within a few 10\,ms when switching off the magnetic potential, due to a weak remaining anticonfinement.

\begin{figure}[tbp]
\begin{center}
\includegraphics[width=3.3in]{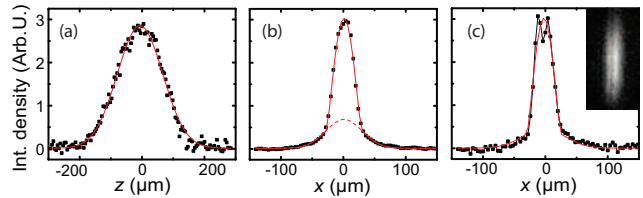}
\caption{Quantum gas deep in 2D regime. Time of flight images from standing wave trap: (a) the vertical density profile is Gaussian, showing that the system is deep in the 2D configuration. In the lateral direction, it is well described by a bimodal Thomas-Fermi profile with $\approx 20$\% thermal fraction in this case: (b) integrated lateral profile averaged over 86 samples, (c) density fluctuations caused by thermal 2D phase fluctuations in single profile from same data set (inset shows $140\times280$\,$\mu$m$^2$) } \label{figure3}
\end{center}
\end{figure}

The 2D regime manifests itself as a change of the shape of the momentum distribution measured in time of flight along $z$ direction. We obtain a Gaussian momentum profile along the vertical (Fig. \ref{figure3}a) corresponding to the harmonic confinement of the trap, while the profile along the other direction remains Thomas-Fermi (Fig. \ref{figure3}b). The 2D system with $\approx2\times10^4$ atoms is below the expected BKT transition temperature of $T_{BKT}\approx40$\,nK at which vortices proliferate in the condensate \cite{Bloch08}. 
Below $T_{BKT}$, phase fluctuations in the condensate fraction can be present which are mapped to density fluctuations \cite{Dettmer01} in time of flight as seen in Fig. \ref{figure3}c. These fluctuations decrease with the temperature of the condensate during the second evaporation step. Apart from the surface evaporation method discussed earlier, the SW configuration allows us to perform forced RF evaporation inside the 2D trap using a radial magnetic field gradient to evaporate along the outer edge of the trap volume.
By stabilizing the bias magnetic field of the QUIC trap to $\sim 3$\,ppm, we obtain a stable evaporation process with reproducible atom numbers in the desired range of several $10^4$ atoms.

The $1/e$ lifetime in the SW trap is $7.8\pm0.4$\,s, and we have used both the incoherent light and a narrow band 765\,nm CW Ti:Sapphire laser for comparison. We find the lifetimes to be the same within the error of the measurements and consistent with loss due to spontaneous emission, indicating that there are no additional loss/heating processes (e.g. photoassociation) associated with using the broadband light source.

\begin{figure}[htbp]
\begin{center}
\includegraphics[width=3.2in]{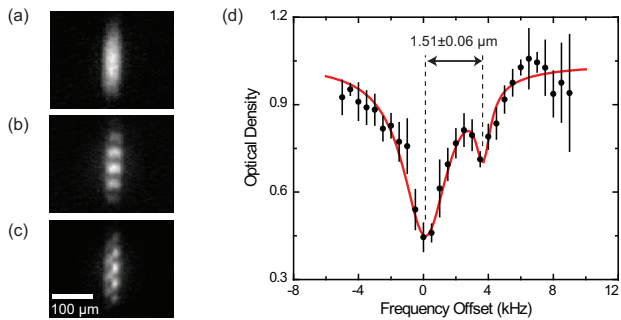}
\caption{Loading of single and multiple planes in the standing wave trap. (a) atoms loaded into a single site, imaged from the side after release and 17\,ms time-of-flight, (b) interference between atoms from two planes. (c) interference pattern between two planes at a higher temperature showing the presence of a thermally activated vortex. (d) The vertical density profile obtained by RF addressing shows the occupation of two planes. The red line denotes a two-peak Lorentzian fit.} \label{figure4}
\end{center}
\end{figure}

We can also intentionally load two planes (here the second and third) of the SW trap by weakening the confinement and choosing a larger separation from the surface in the initial EW configuration.
When two planes are loaded, a vertical sinusoidal interference pattern appears after ballistic time of flight (Figure \ref{figure4}), which can be used to detect long-range phase fluctuations and vortices in the trap \cite{Stock05}.
In order to quantitatively probe the distribution over the trap nodes we employ RF spectroscopy similar to that done in \cite{Stock05}, using a magnetic field gradient of $33.8\pm0.7$\,G/cm perpendicular to the surface. The density distribution along the gradient direction is then probed by varying the RF frequency according to a scaling of $2.43\pm0.05$\,kHz$/\mu$m. This achieves a spatial resolution better than 1\,$\mu$m. The vertical potential periodicity is $1.54\pm0.04$\,$\mu$m, determined by diffracting atoms off the SW.  
The profile is shown in Fig. \ref{figure4}d using an RF pulse length of 2\,s. 
The two loaded sites can be clearly distinguished. Conversely, when loading a single site, the profiling yields an upper limit for the occupation in the second site of $\approx 5$\%, while the (lack of) interference during ballistic expansion limits the fraction of the total coherent population in that site to less than $\approx 10^{-3}$. 

In conclusion, we demonstrate a novel scheme of creating a deeply 2D quantum gas close to a glass surface. The trap provides both strong confinement in the vertical direction and smooth potentials in the 2D plane without the necessity to load many planes simultaneously. We avoid interference of scattered light with the trap light by employing light sources which have short coherence lengths. We create 2D BECs and are able to detect properties such as phase fluctuations and thermal excitation of vortices. By making the trapping surface a part of a microscope, we achieve a very high aperture optical access that should enable resolutions of 500\,nm (RMS) for imaging, creating optical potentials and manipulating atoms. These capabilities enable a wide range of experiments, including optical lattices that combine single-site resolution and significant tunnel coupling in the ground state and experiments on phase fluctuations and vortices in 2D quantum gases.

\begin{acknowledgments}
We acknowledge support from NSF, AFOSR, DARPA and Sloan. J.I.G. and S.F. acknowledge additional support from the NSF and IQSE, respectively.
\end{acknowledgments}

\textsl{Note:} While completing this manuscript, we learned of work describing another approach of combining magnetic and EW potentials, designed for quantum optics \cite{bender08}.

\end{document}